\documentstyle[aps,prl,cite,enumerate,verbatim,epsfig]{revtex}
\begin{document}
\title{Greenberger-Horne-Zeilinger paradoxes for $N$ qu$N$its }
\author{Dagomir Kaszlikowski$^{1,2}$,
Marek \.Zukowski$^{3}$ }
\address{$^1$Department of Physics, National University of Singapore, 10
Kent Ridge Crescent, Singapore 119260 \\
$^2$Instytut Fizyki Do\'swiadczalnej, Uniwersytet Gda\'nski, PL-80-952, Gda\'nsk, Poland,\\
$^5$Instytut Fizyki
Teoretycznej i Astrofizyki, Uniwersytet Gda\'nski, PL-80-952,
Gda\'nsk, Poland.}

\maketitle

\begin{abstract}
In this paper we show the series of Greenberger-Horne-Zeilinger paradoxes for
$N$ maximally entangled $N$-dimensional quantum systems.
\end{abstract}

\section{Introduction}
The Greenberger-Horne-Zeilinger (GHZ) correlations, discovered in 1989,
started a new chapter in the research related to entanglement. To
a great extent this discovery was responsible for the sudden renewal
in the interest in this field, both in theory and experiment.
All these developments finally led to the first actual observation
of three qubit GHZ correlations in 1999 \cite{INNSBRUCK},
and as a by product, since
the experimental techniques involved were of the same kind,
to the famous teleportation experiment \cite{BOUW}.

In \cite{GHZZK,CERF} it has been shown that GHZ paradox can be extended
to the correlations observed in quantum systems consisting of
$N+1$ maximally entangled $N$-dimensional quantum objects (so called qu$N$its), where
$N$ is an arbitrarily high integer number. The existence of the GHZ paradox for
three maximally entangled three dimensional quantum systems has been shown
in \cite{BOOK}. It is worth mentioning that two different approaches have been
presented in \cite{GHZZK,CERF} to derive GHZ paradoxes. In \cite{GHZZK}
the series of paradoxes has been derived for the correlation function that
results from the correlations observed in so called
unbiased symmetric multiport beamsplitters
(for the description of such devices see below) whereas \cite{CERF}
presents derivation based on the relations between operators.

In this paper we would like to show that GHZ-type paradoxes exist
also in the case of correlations expected in gedanken experiments involving
$N$ maximally entangled qu$N$its. This is an important question as the existence
of such paradoxes allows us to deeper understand the structure of
quantum entanglement, which is the fundamental resource in quantum information
as well as the nature of non-classicality of quantum correlations (by non-classicality
we understand lack of local realistic description of such correlations; the alternative
measure of non-classicality can be a measure based on the notion of non-separability).

To this end, we shall study a
GHZ-Bell type experiment in which one has a source emitting
$N$ qu$N$its in a specific  entangled state of the property, that the
qu$N$its propagate towards one of $N$ spatially separated
non conventional measuring devices operated by independent observers.
Each of the devices consists of an unbiased symmetric multiport beam splitter
\cite{ZZH} (with $N$ input and $N$ exit ports), $N$ phase shifters
operated by the observers (one in front of each input), and $N$
detectors (one behind each exit port).

\section{Unbiased multiport beamsplitters}
An unbiased symmetric $2N$-port beam splitter is defined as an $N$-input and
$N$-output interferometric device which has the property that a beam
of light entering via single port is evenly split between all output
ports. I.e., the unitary matrix defining such a device has the
property that
the modulus of all its elements equals ${1\over \sqrt N}$.


An extended
introduction to the physics and theory of such devices is given in
\cite{ZZH}. Multiport beam splitters were introduced into the literature on the EPR paradox
in \cite{KLYSHKO88,ZBGHZ93,ZZHBG94} in order to extend two
qubit Bell-phenomena to observables described as operators in Hilbert spaces of the dimension
higher than two. In contradistinction to the higher than $1/2$ spin
generalizations of the Bell-phenomena \cite{MERMIN80,GARG82,MERMIN82,ARDEHALI91,AGARWAL93,WODKIEWICZ94,WODKIEWICZ95},
this type of
experimental devices generalize the idea of beam-entanglement
\cite{HORNE85,ZUKOWSKI88,HORNE89,RARITY90}.
Unbiased symmetric multiport beam splitters are performing unitary
transformations between "mutually unbiased" bases in the Hilbert space
\cite{SCHWINGER60,IVANOVIC81,WOOTERS86}.
They were tested in several recent experiments
\cite{MMWZZ95,RECKPHD96}, and also various aspects of such devices were
analyzed theoretically \cite{RECK94,JEX95}.

We shall use here only multiport beam splitters which have the property
that the elements of the unitary transformation which describes their
action are given by
\begin{equation}
U_{m,m'}^{N}={1\over \sqrt N}\gamma_{N}^{(m-1)(m'-1)},
\end{equation}
where $\gamma_{N}=\exp(i{2\pi\over N})$ and the indices $m$, $m'$
denote the input and exit ports. Such devices were called in
\cite{ZZH} the Bell multiports.

\section{Quantum mechanical predictions}

Although in this paper we consider only the situation in which there is
an equal number of observers and input ports in each multiport beamsplitter it
is instructive to derive necessary formulas for the more general case in which there
are $M$ observers each operating $2N$ port beamsplitter.
The initial $M$ qu$N$it state has the following form:

\begin{eqnarray}
\label {eq1}
&&|\psi(M)\rangle={1\over \sqrt
N}\sum_{m=1}^{N}\prod_{l=1}^{M}|m\rangle_{l},
\label{2}
\end{eqnarray}
where $|m\rangle_{l}$ describes the $l$-th qu$N$it being in the
$m$-th beam, which leads to the $m$-th input of the $l$-th multiport.
Please note, that only one qu$N$it enters each multiport.
However, each of the qu$N$its itself is in a mixed state (with equal
weights), which gives it equal probability to enter the local
multiport via any of the input ports.


The state (\ref{2}) seems to be the most straightforward
generalization of the GHZ states to the new type of observables. In
the original GHZ states the number of their components (i.e., two) is
equal to the dimension of the Hilbert space describing the relevant
(dichotomic) degrees of freedom of each of the qu$N$its. This
property is shared with the EPR-type states proposed in \cite{ZZH} for
a two-multiport Bell-type experiment. In this case the number of
components equals the number of input ports of each of the multiport
beam splitters. We shall not discuss here the possible methods to
generate such states. However, we briefly mention that the recently
tested entanglement swapping \cite{EVENT,ZZW,Pan98}
technique could be used for this purpose.

As it was mentioned earlier, in front of every input of each multiport
beam splitter one has a tunable phase shifter. The initial state is
transformed by the phase shifters into

\begin{eqnarray}
\label {eq2}
&&|\psi(M)'\rangle={1\over \sqrt N}\sum_{m=1}^{N}\prod_{l=1}^{M}
\exp(i\phi_{l}^{m})|m\rangle_{l},
\end{eqnarray}
where $\phi_{l}^{m}$ stands for the setting of the phase shifter in
front of the $m$-th port of the $l$-th multiport.

The quantum prediction for  probability to register the first photon
in the output $k_{1}$ of an $2N$ - port device, the second one in the
output $k_{2}$ of the second such device ,$\dots$, and the $M$-th one in
the output $k_{M}$ of the $M$-th device is given by:
\begin{eqnarray}
 &P_{QM}(k_{1},\dots,k_{M}|\vec{\phi_{1}},\dots,\vec{\phi_{M}})=&\nonumber\\
 &({1\over N})^{M+1}|\sum_{m=1}^{N}\exp(i\sum_{l=1}^{M}\phi_{l}^{m})
 \prod_{n=1}^{M}\gamma_{N}^{(m-1)(k_{n}-1)}|^{2}=&\nonumber\\
 &=({1\over N})^{M+1}\left[N+2\sum_{m>m'}^{N}
 \cos\left(\sum_{l=1}^{M}\Delta\Phi_{l,k_{l}}^{m,m'}
 \right)\right],&
\label{5}
\end{eqnarray}
where
$\Delta\Phi_{l,k_{l}}^{m,m'}=\phi_{l}^{m}-\phi_{l}^{m'}+{2\pi\over
N}(k_{l}-1)(m-m')$. The shorthand symbol
$\vec{\phi}_{k}$ stands for the full set of phase settings in front of
the $k$-th multiport, i.e.
$\phi_{k}^{1},\phi_{k}^{2},\dots,\phi_{k}^{N}$.

To efficiently describe the local detection events
let us employ a specific value assignment method (called Bell
number assignment; for a detailed
explanation see again \cite{ZZH}), which ascribes to the detection event
behind the $m$-th output of a multiport the value  $\gamma_{N}^{m-1}$, where
$\gamma_{N}=\exp(i{2\pi\over N})$.
With such a value assignment to the detection events, the Bell-type
correlation function, which is the average of the product of the
expected results, is defined as

\begin{eqnarray}
&E(\vec{\phi_{1}},\cdots,\vec{\phi_{M}})=&\nonumber\\
&=\sum_{k_{1},\cdots,k_{M}=1}^{N}
\prod_{l=1}^{M}\gamma_{N}^{k_{l}-1}P(k_{1},\cdots,k_{M}|\vec{\phi_{1}},\cdots,\vec{\phi_{M}})&
\label{eq4}
\end{eqnarray}
and as we shall see for the quantum case it acquires particularly simple
and universal form (which is the main purpose for using this
non-conventional value assignment).

The easiest way to compute the correlation function for the quantum
prediction employs the mid formula of (\ref{5}):
\begin{eqnarray}
&E_{QM}(\vec{\phi_{1}},\cdots,\vec{\phi_{M}})=&\nonumber\\ &=({1\over
N})^{M+1}\sum_{k_{1},\cdots,k_{M}=1}^{N}\sum_{m,m'=1}^{N}
\exp\left(i\sum_{n=1}^{M}(\phi_{n}^{m}-\phi_{n}^{m'})\right)&\nonumber\\
&\times \prod_{l=1}^{M}\gamma_{N}
^{(k_{l}-1)(m-m'+1)}=&\nonumber\\
&=({1\over N})^{M+1}\sum_{m,m'=1}^{N}
\exp\left(i\sum_{n=1}^{N}(\phi_{n}^{m}-\phi_{n}^{m'})\right)&\nonumber\\
&\times\prod_{l=1}^{M}
\sum_{k_{l}=1}^{N}\gamma_{N}
^{(k_{l}-1)(m-m'+1)}.&
\label{corr1}
\end{eqnarray}
Now, one notices that
$\sum_{k_{l}=1}^{N}\gamma_{N}^{(k_{l}-1)(m-m'+1)}$
differs from zero
(and equals to N)
only if $m-m'+1=0$ modulo N. Therefore we can finally write:

\begin{eqnarray}
&E_{QM}(\vec{\phi}_{1},\cdots,\vec{\phi}_{M})\nonumber&\\
&={1\over N}\sum_{m=1}^{N}\exp(i\sum_{l=1}^{M}\phi^{m,m+1}_{l}),&
\label{corr2}
\end{eqnarray}
where $\phi^{m,m+1}_{l}=\phi^{m}_{l}-\phi^{m+1}_{l}$ and the above
sum is modulo N, i.e.,
$\phi^{N+1}_{l}=\phi^{1}_{l}$.

One can notice here a striking simplicity and symmetry of this quantum
correlation function (\ref{corr2}). It is valid for all possible
values of M (number of qu$N$its) and for all possible values of
$N\geq 2$ (number of ports). For $N=2$, it reduces itself to the usual
two qubit, and for $N=2$, $M\geq2$ the standard GHZ type
multi-qubit correlation function for beam-entanglement experiments,
namely $\cos(\sum_{l=1}^{M}\phi_{l}^{1,2})$ \cite{MERMIN90}. The Bell-EPR phenomena discussed in
\cite{ZZH} are described by (\ref{corr2}) for $M=2, N\geq 3$.

Even for $N=2$, $M=1$ the function (\ref{corr2}) describes
the following process. Assume that
a traditional four-port 50-50 beam splitter, is fed a single photon
input in  a state in which is an equal superposition of being in each
of the two input ports. The value of (\ref{corr2}) is the average of
expected photo counts behind the exit ports (provided the click at one
of the detectors is described as $+1$ and at the other one as $-1$),
and of course it depends on the relative phase shifts in front of the
beam splitter. In other words, this situation describes a Mach-Zehnder
interferometer with a single photon input at a chosen input port. For
$N=3$, $M=1$ the same interpretation applies to the case of a
generalised three input, three output Mach-Zehnder interferometer
described in \cite{WEIHS96}, provided one ascribes to firings of
the three detectors respectively
$\gamma_{3}=\alpha\equiv\exp(i{2\pi\over 3})$, $\alpha^{2}$ and
$\alpha^{3}$.

The described set of gedanken experiments is rich in EPR-GHZ
correlations (for $M\geq2$). To reveal the above, let us first analyze
the conditions (i.e. settings) for such correlations. As the
correlation function (\ref{corr2}) is an average of complex numbers of
unit modulus, one has
$|E_{QM}(\vec{\phi}_{1},\cdots,\vec{\phi}_{M})|\leq 1$. The equality
signals a perfect EPR-GHZ correlation. It is easy to notice that this
may happen only if
$$\exp(i\sum_{l=1}^{M}\phi_{l}^{1,2})=
\exp(i\sum_{l=1}^{M}\phi_{l}^{2,3})=\cdots=
\exp(i\sum_{l=1}^{M}\phi_{l}^{M,1})=\gamma_{N}^{k},$$ where k is an arbitrary
natural number. Under this condition
$E(\vec{\phi}_{1},\cdots,\vec{\phi}_{M})=\gamma_{N}^{k}$. This means that only
those sets of $M$ spatially separated detectors may fire, which are
ascribed such Bell numbers having the property that their product
is $\gamma_{N}^{k}$. Knowing, which detectors fired in the set of
$M-1$ observation stations, one can predict with certainty which
detector would fire at the sole observation station not in the set.

\section{GHZ paradoxes for $N$ maximally entangled qu$N$its.}

Now we show the derivation of GHZ paradoxes for $N$ maximally entangled
qu$N$its. However, first it is instructive to consider the simplest case, which
is three qutrits.
The correlation function for such
an experiment reads
\begin{eqnarray}
&&E_{QM}(\vec{\phi}_{1},\vec{\phi}_{2},\vec{\phi}_{3})\nonumber\\
&&={1\over 3}\sum_{k=1}^{3}
\exp\left(i\sum_{l=1}^{3}(\phi_{l}^{k}-
\phi^{k+1}_{l})\right).
\label{3trit}
\end{eqnarray}

The experimenters perform three distinctive experiments.
In the first run of the experiment we allow the observers
to choose the following settings of the measuring apparatus,
$\vec{\phi}_{1}=\vec{\phi}_{2}=(0,{\pi\over 3},{2\pi\over3})=\vec{\phi},
\vec{\phi}_{3}=(0,0,0)=\vec{\phi}'$
in the second run they choose
$\vec{\phi}_{2}=\vec{\phi}_{3}=\vec{\phi}, \vec{\phi}_{1}=\vec{\phi}'$ whereas in the
third run they fix the local settings of their tritters on
$\vec{\phi}_{1}=\vec{\phi}_{3}=\vec{\phi}, \vec{\phi}_{2}=\vec{\phi}'.$

Now, let us calculate the numerical values of
the correlation function for each experimental
situation. We easily find that for all three
experiments this value, due to the special form
of the correlation function, is the same and reads
\begin{eqnarray}
&&E_{QM}(\vec{\phi},\vec{\phi},\vec{\phi}')=
E_{QM}(\vec{\phi}',\vec{\phi},\vec{\phi})=
E_{QM}(\vec{\phi},\vec{\phi}',\vec{\phi})\nonumber\\
&&=\exp(-i{2\pi\over 3})=\alpha^2,
\end{eqnarray}
i.e., we observe perfect correlations.

The (deterministic) local hidden variable correlation
function for this type of experiment must have the following structure
\cite{Bell64}:

\begin{eqnarray}
&&E_{HV}(\vec{\phi}_{1},\vec{\phi}_{2},\vec{\phi}_{3})
=\int_{\Lambda}\prod_{k=1}^{3}I_{k}(\vec{\phi}_{k},\lambda)\rho(\lambda)d\lambda.
\end{eqnarray}
The hidden variable function $I_{k}(\vec{\phi}_{k},\lambda)$, which
determines the firing of the detectors behind the k-th multiport,
depends only upon the local set of phases, and takes one of the three
possible values $\alpha$, $\alpha^{2}$, $\alpha^{3}=1$ (these values
indicate which of the detectors is to fire), and $\rho(\lambda)$ is
the distribution of hidden variables. The local realistic
description of the experiment is only possible if the local hidden variable
correlation function defined above equals that of quantum ones for the given set
of phase shifts.

Because in each experiment described above we observe perfect correlations, i.e.,
the quantum mechanical correlation function takes value $\alpha^2$, the local realistic
description is possible if and only if one has $E_{HV}(\vec{\phi},\vec{\phi},\vec{\phi}')=
E_{HV}(\vec{\phi}',\vec{\phi},\vec{\phi})=
E_{HV}(\vec{\phi},\vec{\phi}',\vec{\phi})=\alpha^2$. Taking into account the structure of
local hidden variable correlation function one obtains the set of three equations

\begin{eqnarray}
&&I_{1}(\vec{\phi},\lambda)I_{2}(\vec{\phi},\lambda)I_{3}(\vec{\phi}'
,\lambda)=
\alpha^{2}\nonumber\\
&&I_{1}(\vec{\phi}',\lambda)I_{2}(\vec{\phi},\lambda)I_{3}(\vec{\phi}
,\lambda)=
\alpha^{2}\nonumber\\
&&I_{1}(\vec{\phi},\lambda)I_{2}(\vec{\phi}',\lambda)I_{3}(\vec{\phi}
,\lambda)=
\alpha^{2}
\label{telephone1}
\end{eqnarray}
After multiplication of the above equations one arrives at:
\begin{eqnarray}
&&\prod_{k=1}^{3}I_{k}(\vec{\phi}',\lambda)\prod_{k=1}^{3}I_{k}(\vec{\phi},
\lambda)^{2}=(\alpha^{2})^{3}=1,
\label{telephone2}
\end{eqnarray}
which can be also written in the following form
\begin{eqnarray}
&&\prod_{k=1}^{3}I_{k}(\vec{\phi}',\lambda)=\prod_{k=1}^{3}I_{k}(\vec{\phi},
\lambda).
\label{telephone3}
\end{eqnarray}
We have used the property of the hidden variable functions:
$I_{k}(\vec{\phi},\lambda)^{2}=I_{k}(\vec{\phi},\lambda)^{*}$ for
every $\lambda$. However, because $E_{QM}(\vec{\phi}',\vec{\phi}',
\vec{\phi}')=1$ we must also have
\begin{eqnarray}
&&\prod_{k=1}^{3}I_{k}(\vec{\phi}',
\lambda)=1,
\label{maja}
\end{eqnarray}
which, because of (\ref{telephone3}), gives
\begin{eqnarray}
&&\prod_{k=1}^{3}I_{k}(\vec{\phi},
\lambda)=1
\label{maja2}
\end{eqnarray}
for every $\lambda$. This in turn implies that
\begin{eqnarray}
&&E_{QM}(\vec{\phi},\vec{\phi},
\vec{\phi})=1,
\end{eqnarray}
which means that local hidden variables predict perfect correlations for
the experiment when all observers set their local settings at
$\vec{\phi}$. However, the quantum prediction is that
\begin{eqnarray}
&&E_{QM}(\vec{\phi},\vec{\phi},\vec{\phi})=-{1\over 3}.
\label{Maja3}
\end{eqnarray}
Therefore, we have the contradiction: $1=-{1\over 3}$.

This contradiction is of the different
type than the one derived in \cite{GHZZK} although it has been
obtained in the similar way, i.e., the perfect correlations have been
used to derive it (equations (\ref{telephone1}) and (\ref{telephone2})).
Here local hidden variables
imply a certain perfect correlation, which is not predicted by quantum
mechanics, whereas in \cite{GHZZK} as well as in \cite{CERF} the contradiction is that both
theories predict perfect correlations but of the different type.

Now, we employ the above procedure for the case when we have an arbitrary odd
number of multiports and qu$N$its $N=2m+1$. As we have seen in
in the above derivation the crucial point is to find the
proper phases for the multiports. Let us choose for the
first gedanken experiment the following ones
$\vec{\phi}_{1}=\vec{\phi}_{2}=\vec{\phi}_{3}=...=\vec{\phi}_{2m}=
(0,{\pi\over 2m+1},{2\pi\over 2m+1},\dots,{2m\pi\over 2m+1})=\vec{\phi},
\vec{\phi}_{2m+1}=(0,0,...,0)=\vec{\phi}'$.

As before, in the next run of the experiment we choose the same
phases but we change the role of observers such that in the second run
the first
one chooses $\vec{\phi}'$ while the rest of them choose $\vec{\phi}$,
in the third run the third one chooses $\vec{\phi'}$ while the rest
of them choose
$\vec{\phi}$, etc. Again, the value of the correlation
function for each experiment is the same
\begin{eqnarray}
&&E_{QM}(\vec{\phi},\dots,\vec{\phi}')=
E_{QM}(\vec{\phi}',\dots,\vec{\phi})=\dots=
E_{QM}(\vec{\phi},\dots,\vec{\phi}',\vec{\phi})\nonumber\\
&&={1\over 2m+1}\left[2m\exp\left(-i{2m\pi\over 2m+1}\right)+\exp\left(i{4m^{2}\pi\over
2m+1}\right)\right]=\gamma_N^{-m}
\label{vdots}
\end{eqnarray}
Using (\ref{vdots}) and the structure of the hidden
variables correlation function, i.e., $E_{HV}(\vec{\phi}_1,\dots,\vec{\phi}_N)=
\int_{\Lambda}\prod_{k=1}^{N}I_{k}(\vec{\phi}_{k},\lambda)\rho(\lambda)d\lambda$,
we arrive at the set of $N$ equations
$I_{l}(\vec{\phi}',\lambda)\prod_{k\neq l}^{N-1}I_{k}(\vec{\phi},\lambda)=\gamma_N^{-m}$ ($l=1,2,\dots, N$). After
multiplying them by each other we obtain
\begin{eqnarray}
&&\prod_{k=1}^{2m+1}I_{k}(\vec{\phi}',\lambda)=
\prod_{k=1}^{2m+1}I_{k}(\vec{\phi},\lambda)^{-2m}=
\prod_{k=1}^{2m+1}I_{k}(\vec{\phi},\lambda)
\label{singapore}
\end{eqnarray}
for every $\lambda$. Because $E_{QM}(\vec{\phi}',\dots,\vec{\phi}')=1$
one must have
$\prod_{k=1}^{2m+1}I_{k}(\vec{\phi}',\lambda)=1$,
which according to (\ref{singapore}) gives
$\prod_{k=1}^{2m+1}I_{k}(\vec{\phi},\lambda)=1$
for every $\lambda$. Thus, local hidden variables imply
the following perfect correlation
$E_{QM}(\vec{\phi},\dots,\vec{\phi})=1$,
whereas quantum mechanics gives
\begin{eqnarray}
&&E_{QM}(\vec{\phi},...,\vec{\phi})
={1\over 2m+1}\left[2m\exp\left(-i\frac{2m+1}{2m+1}\pi\right)\right.
\left.+\exp\left(i\frac{(2m+1)2m}{2m+1}\pi\right)\right]={-2m+1\over 2m+1}.
\end{eqnarray}
Therefore, for each $m$ one obtains
the untrue identity $1=\frac{-2m+1}{2m+1}$,
which in the limit of $m\longrightarrow \infty$ becomes
$1=-1$.

For the even number of qu$N$its
and the multiports $N=2m$ ($m\geq 2$) we choose the following
phases:
$\vec{\phi}_{1}=\vec{\phi}_{2}=\vec{\phi}_{3}=\dots
=\vec{\phi}_{N-1}=(0,{\pi\over 2m-1},{2\pi\over 2m-1},\dots,{(2m-1)\pi\over 2m-1})=\vec{\phi},
\vec{\phi}_{N}=(0,0,...,0)=\vec{\phi'}.$
One easily finds that
\begin{eqnarray}
&&E_{QM}(\vec{\phi},\dots,\vec{\phi}')=E_{QM}(\vec{\phi}',\dots,\vec{\phi})
=\dots=E_{QM}(\vec{\phi},\dots,\vec{\phi}',\vec{\phi})\nonumber\\
&&={1\over 2m}\left[(2m-1)\exp\left(-i\frac{2m-1}{2m-1}\pi\right)\right.
\left.+\exp\left(i\frac{(2m-1)^{2}}
{2m-1}\pi\right)\right]=-1
\end{eqnarray}
and
\begin{eqnarray}
&&E_{QM}(\vec{\phi},..,\vec{\phi})\nonumber\\
&&={1\over 2m}
\left[(2m-1)\exp\left(-i\frac{2m}{2m-1}\pi\right)+
\exp\left(i\frac{2m(2m+1)}{2m+1}\pi\right)\right]\nonumber\\
&&={1\over 2m}\left[(2m-1)\exp\left(-i\frac{2m}{2m-1}\pi\right)+1\right].
\end{eqnarray}
Applying the same reasoning as above
one has $1={1\over 2m}\left[(2m-1)\exp\left(-i\frac{2m}{2m-1}\pi\right)+1\right].$
Again, for each $m>2$ one has a contradiction, which
in the limit of $m\longrightarrow\infty$ becomes $1=-1$.


\section{Conclusions}

We have derived the series of paradoxes in which the number of
observers and the dimension of the Hilbert space describing each subsystem
is the same.
The derivation of the paradoxes relies on the perfect
correlations observed in the system but the resulting contradiction between
quantum mechanics and local realism
manifests itself in the fact that local realism predicts a certain perfect
correlation whereas quantum mechanics does not.

An interesting feature of the paradoxes is that they naturally
split into two parts: the even and the odd number of observers (input ports). In each
part the final contradiction has different numerical values. All paradoxes do
not vanish with the growing dimension $N$ of the Hilbert space
of each qu$N$it.

Furthermore, for the case of two observers one cannot derive the GHZ
paradoxes with the method presented here (for this case we have so called Hardy paradox but for
non-maximally entangled state \cite{HARDY}).

In conclusion we state that the multiport beam splitters, and the idea
of value assignment based Bell numbers, lead to a strikingly
straightforward generalisation of the GHZ paradox for $N$ entangled
qu$N$its.
These properties may possibly find an application in
future quantum information and communication schemes (especially as
GHZ states are now observable in the lab \cite{INNSBRUCK}).
\\
\\
{\bf Acknowledgements}
\\
\\
MZ and DK are supported by the KBN grant No. 5 P03B 088 20.

\end{document}